# Compilation of Hazardous and Benign Material Complex Dielectric Constants at Millimeter Wave Frequencies for Security Applications

Mahshid Asri[1], Elizabeth Wig[2], Carey Rappaport[1]

[1] Department of Electrical and Computer Engineering, Northeastern University, Boston, MA, USA, asri.m@northeastern.edu

[2] Department of Electrical Engineering, Stanford University, Stanford, CA, USA

*Abstract*—This paper compiles measured complex dielectric constants of benign and hazardous materials used for developing algorithms for personnel scanners at airports at 30 GHz. The materials are grouped into broad classifications of potential threat by mapping the materials to the complex dielectric constant plane.

## I. Introduction

In order to improve efficiency in airport security person scanning, it is necessary to upgrade the algorithms that discriminate possible threats. Several studies [1-3] have discussed developing automatic algorithms that can distinguish benign objects from threats. Fast classification of body-worn objects into potential threats and safe objects would minimize pat-downs and shorten security screening time.

Previously developed algorithms [1,2] can automatically predict the permittivity and thickness of attached objects. To classify the person-worn material, it is necessary to define some criteria on the predicted permittivity to indicate the threatening level of the object. Here, a set of boundaries on the complex permittivity of predicted materials is presented which is used to classify objects into threat and non-threat categories. A collection of typical materials was identified, and their measured dielectric constants was found in the literature. These permittivities are compiled in Table I and plotted on the complex permittivity plane.

## II. Permittivity Plane

In this section, the complex dielectric constant of materials commonly used in the development of airport security algorithms are presented. The most popular center frequency of wideband mm-wave Advanced Imaging Technology (AIT) scanners, both in the field and being newly developed, is 30 GHz. The permittivities of typically concealed materials that are presented in Table I were measured at 30 GHz or extrapolated from 5 to 20 GHz to 30 GHz.

Table I. Complex permittivity of typical materials at 30 GHz.

| Material | Complex Permittivity | Material | Complex Permittivity |
|---|---|---|---|
| Silicone rubber [5] | 3 - j 0.001 | Flour [9] | 1.9 - j 0.075 |
| Petroleum Jelly [6] | 2.15 - j 0.0007 | Soap [9] | 2.75 - j 0.225 |
| Jujube Honey [7] | 8.7 - j 4.8 | Wood [9] | 2.55 - j 0.14 |
| Baking Soda [8] | 2.5 - j 0.025 | Salt [9] | 3.05 - j 0.015 |
| Sugar [9] | 3.5 - j 0.0025 | Sand 1.9 gr/cm$^3$ [10] | 4.5 – j 0.04 |
| Powdered Sugar [10] | 2.05 - 0.004 | Sand 1.8 gr/cm$^3$ [10] | 5.9 - j 0.01 |
| Talc [8] | 1.75 - j 0.01 | Plexiglass [15] | 2.51 - j 0.01 |
| Sheet glass (heated to 1737 F) [11] | 5.29 - j 0.125 | Glass, High Purity Fused Silica [11] | 3.75 - j 0.0035 |
| Denim [12] | 1.6 - j 0.015 | Red Leather [12] | 2.2 - j .09 |
| TNT [13] | 2.84 - j 0.005 | PETN [13] | 2.38 - j 0.02 |
| RDX [13] | 2.60 - j .01 | C4 [13] | 3.28 - j 0.04 |
| Cocaine [13] | 3 - j 0.01 | Ethanol [14] | 4.5 - j 1.5 |
| Methanol 0.6 Mol Solution [14] | 7 - j 7 | Water [16] | 20 – j 30 |
| Paper [9] | 2.35 - j .11 | Dry Skin [17] | 20 – j16 |

Figure 1 shows a scatter plot of collected materials on the permittivity plane. As shown in the lower red box of Figure 1, it is possible to define an area on the permittivity plane that includes threats like TNT, PETN, RDX, and C4. Although there are some materials in the lower red box that can be assumed benign such as sugar, salt, and baking soda, because of their close permittivity to the unsafe materials they can be considered threat surrogates. Subjects carrying them would be sent to secondary checking. For all the materials in the lower red box, the object's back surface can be seen on radar images, and they are considered practically lossless, meaning their loss tangent cannot be calculated from information available on their reconstructed images. The green box shows an area in which all the materials are considered safe. The upper right red box includes possible peroxides and similar water-based

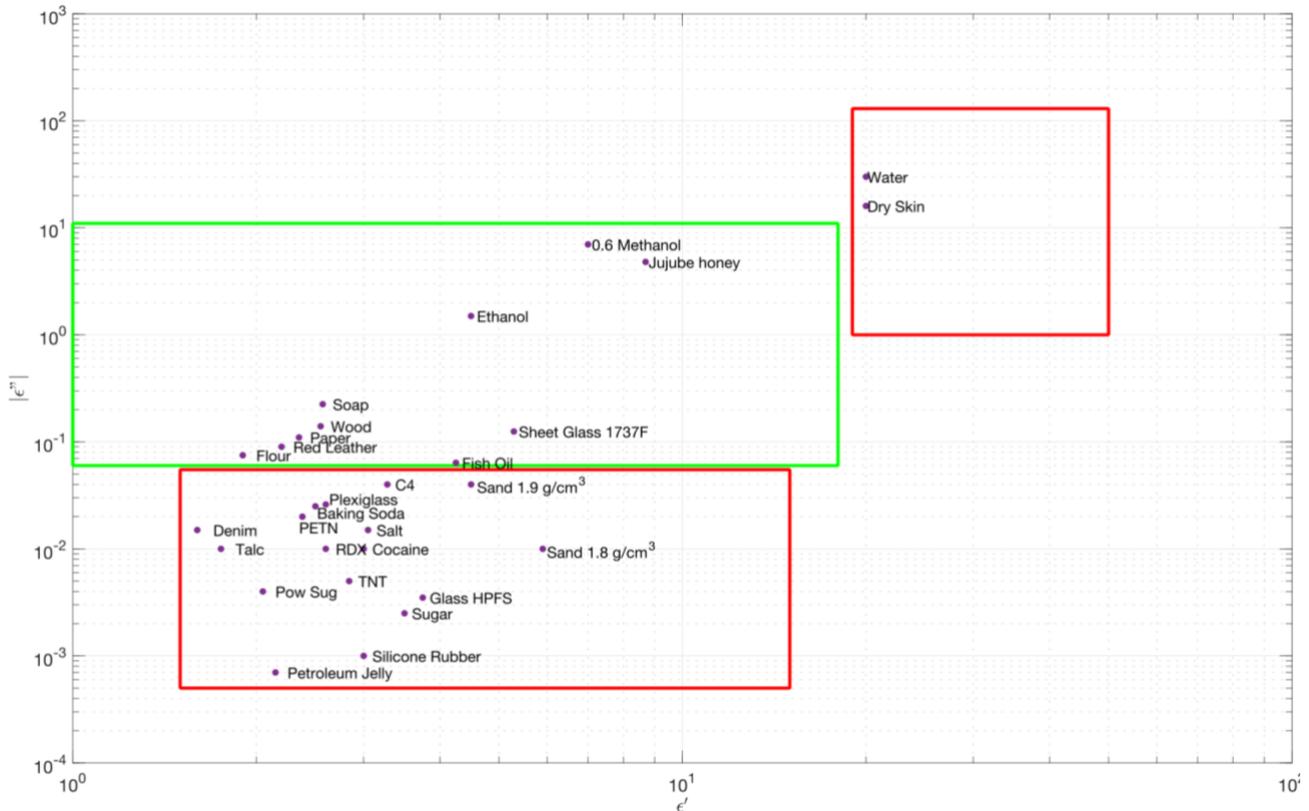

Fig. 1. Log-log plot of complex permittivities at 30 GHz.

materials, with surrogates like dish soap and body lotion also falling in the upper red box.

Having these criteria for the dielectric constant, it is possible to predict the threatening level of a person-worn object. If no back surface is detected on the image, the object is considered lossless and the characterization algorithm explained in [2] only predicts the real part of the permittivity. Since the lower red box indicates the lossless materials, the imaginary part of an object predicted lossless can be considered to be between 0.0005 and 0.055. By plotting the predicted complex permittivity and observing the box in which it falls, the algorithm can decide if the object is a potential threat or not.

For slightly lossy materials, only the reflection from the front object surface is visible. Using the relative intensity of the front surface reflectivity to that of the skin, an estimate of the reflection coefficient can be found. This in turn provides a locus of possible complex permittivities, which can be plotted on the permittivity plane. If the entire curve happens to fall within the green box, the object is predicted to be safe. If any part of the curve extends into either of the red boxes, the object is considered a hazard. If the whole curve is located outside the three boxes, the object will also go through the pat-down process, since it could not be accurately detected as a safe material.

## III. CONCLUSION

Airports' personnel scanners can work faster if they are equipped with a software that can automatically exclude benign objects from potential threats. This paper presents, in one place, the complex dielectric constants of many of the materials that might constitute threats, threat surrogates, or benign materials. Previous studies [1-3] have discussed algorithms that can automatically characterize a weak dielectric object by predicting its permittivity and thickness. Here, it has been shown that by plotting the predicted permittivities and comparing it to the red and green boxes, it is possible to predict if the person-worn object is benign. This will be beneficial in lowering the false alarm rate and improving the air travel experience while maintaining its safety.

## ACKNOWLEDGMENT

This work is supported by the U.S. Department of Homeland Security, Science and Technology Directorate, Office of University Programs, under Grant Award 70RSAT18FR0000115. The views and conclusions contained herein are those of the authors and should not be interpreted as necessarily representing the official policies, either expressed or implied, of the U.S. Department of Homeland Security.

## REFERENCES

[1] Asri M, Rappaport C. Automatic threat prediction of body-worn objects for security screening purposes. unpublished.

[2] Asri M, et al. 2020. Automatic permittivity and thickness characterization of body-bome weak dielectric threats using wideband radar. 14th European Conference on Antennas and Propagation (EuCAP). IEEE, pp. 1–4.

[3] Asri M, and Rappaport C. 2019. Automatic permittivity characterization of a weak dielectric attached to human body using wideband radar image processing. IEEE International Symposium on Antennas and Propagation and USNC-URSI Radio Science Meeting. IEEE, pp. 575–576.

[4] Tajdini MM, et al. 2020. Image radar determining the nominal body contour for characterization of concealed person- worn explosives. *14th European Conference on Antennas and Propagation (EuCAP)*. IEEE, pp. 1–4.

[5] Sarkarat M, et al. 2020. High field dielectric properties of clay filled silicone rubber composites. Materials Today Communications 23 :100947.

[6] Naftaly M, et al. 2015. A comparison method for THz measurements using VNA and TDS. 40th International Conference on Infrared, Millimeter, and Terahertz waves (IRMMW-THz), IEEE, pp. 1-2.

[7] Yang M, et al. 2018. Broadband dielectric properties of honey: effect of water content. Journal of Agricultural Science and Technology, 20.4 :685-693.

[8] Nguyen KN, et al. 2006. Millimeter Wave Dielectric Permittivity Measurements of Common Materials. Joint 31st International Conference on Infrared Millimeter Waves and 14th International Conference on Teraherz Electronics, IEEE, pp. 483-483.

[9] Korolev KA, Afsar M. N. 2005. Complex dielectric permittivity measurements of materials in millimeter waves. Joint 30th International Conference on Infrared and Millimeter Waves and 13th International Conference on Terahertz Electronics, Vol. 2, IEEE, pp. 594-595.

[10] Chen S, et al. 2007. Broad-band millimetre wave spectroscopy of common materials. European Microwave Conference, IEEE, pp. 692-695.

[11] Chen S, et al. 2006. Millimeter-wave dielectric permittivity of glasses. Joint 31st International Conference on Infrared Millimeter Waves and 14th International Conference on Teraherz Electronics, IEEE, pp. 406-406.

[12] Harmer SW, et al. 2008. Determination of the complex permittivity of textiles and leather in the 14–40 GHz millimetre-wave band using a free-wave transmittance only method. IET Microwaves, Antennas & Propagation, 2(6):606-14.

[13] Watters DG, et al. 1995. Microwave inspection of luggage for contraband materials using imaging and inverse-scattering algorithms. Journal of Research in Nondestructive Evaluation, 7(2-3):153-68.

[14] Abeyrathne CD, Skafidas E. 2014. Complex permittivity measurements in 1–30 GHz using a MEMS probe. Journal of Microelectromechanical Systems, 24(4):976-81.

[15] Zhekov SS, et al. 2020. Dielectric properties of common building materials for ultrawideband propagation studies [measurements corner]. IEEE Antennas and Propagation Magazine, 4;62(1):72-81.

[16] Meissner T, Wentz FJ. 2004. The complex dielectric constant of pure and sea water from microwave satellite observations. IEEE Transactions on Geoscience and remote Sensing, 42(9):1836-49.

[17] Dry skin permittivity". [Online]. Available: http://niremf.ifac.cnr.it/docs/ DIELECTRIC/AppendixD.html